\documentclass[sigconf,10pt]{acmart}
\usepackage[english]{babel}
\usepackage{blindtext}
\usepackage{booktabs}
\usepackage{pifont}
\usepackage{amssymb}
\usepackage{balance}
\usepackage{graphicx}
\usepackage{enumitem}
\usepackage{amsmath,amssymb}
\usepackage{float}
\usepackage{dblfloatfix}
\usepackage{url}
\usepackage{multirow}
\usepackage{multicol}
\usepackage{xcolor}
\usepackage{mathtools}
\usepackage{hyperref}
\usepackage{enumitem}
\usepackage[super]{nth}
\usepackage[caption=false]{subfig}
\usepackage[font=small,labelfont=bf]{caption}
\usepackage[export]{adjustbox}
\usepackage[linesnumbered,ruled]{algorithm2e}
\usepackage{soul}
\usepackage{rotating}
\hypersetup{
  colorlinks = true, 
  urlcolor   = blue, 
  linkcolor  = blue, 
  citecolor  = red   
}

\setlist[enumerate]{leftmargin=*,topsep=0pt,itemsep=0ex}

\usepackage[normalem]{ulem}

\definecolor{purple}{rgb}{1,0,1}
\definecolor{red}{rgb}{1,0,0}
\definecolor{blue}{rgb}{0,0,1}

\newcommand{\eg}{\textit{e.g., }}
\newcommand{\ie}{\textit{i.e., }}
\newcommand{\etal}{\textit{et al.}}
\newcommand{\myitem}[1]{\noindent\textbf{#1}}
\newcommand{\myitemit}[1]{\noindent\textbf{\textit{#1}}}

\renewcommand\footnotetextcopyrightpermission[1]{} 
\setcopyright{none}

\acmDOI{}

\acmISBN{}

\acmPrice{}

\begin{document}
\title{Validating IP Prefixes and AS-Paths with Blockchains}

\author{Ilias Sfyrakis}
    \affiliation{
        \institution{University of Crete}
        \city{Heraklion, Greece} 
    }
    \email{csd3078@uoc.csd.gr}

\author{Vasileios Kotronis}
    \affiliation{
        \institution{FORTH}
        \city{Heraklion, Greece} 
    }
    \email{vkotronis@ics.forth.gr}
    
\renewcommand{\shortauthors}{Sfyrakis \etal}

\begin{abstract}
Networks (Autonomous Systems-AS) allocate or revoke IP prefixes with the intervention of official Internet resource number authorities, and select and advertise policy-compliant paths towards these prefixes using the inter-domain routing system and its primary enabler, the Border Gateway Protocol (BGP).
Securing BGP has been a long-term objective of several research and industrial efforts during the last decades, that have culminated in the 
Resource Public Key Infrastructure (RPKI) for the cryptographic verification of prefix-to-AS assignments. However, there is still no widely adopted 
solution for securing IP prefixes and the (AS-)paths leading to them; approaches such as BGPsec have seen minuscule deployment. In this work, we design and implement a Blockchain-based system that (i) can be used to validate both of these resource types, (ii) can work passively and does not require any changes in the inter-domain routing system (BGP, RPKI),
and (iii) can be combined with currently available systems for the detection and mitigation of routing attacks. We present early results and insights w.r.t. scalability.
\end{abstract}

\maketitle

    
    
    

\section{Introduction}\label{sec:intro}
\myitem{Internet Routing.} The Internet is composed of $\sim$64k~\cite{huston2019cidr} interconnected networks (Autonomous Systems--AS) that participate in the global Internet routing system.
Official Internet Registries (IRs), with different areas of jurisdiction, form hierarchies (\eg from IANA~\cite{iana2019} to Regional IRs--RIRs~\cite{riperir2019} to National IRs--NIRs~\cite{grlirs2019ripe}) that manage Internet number resources such as AS Numbers (ASNs) and IP prefixes (\ie groups of IP addresses). These authorities assign IP prefixes to ASes, which can in turn lease/delegate these resources to other ASes.
ASes exchange prefix reachability information via the Border Gateway Protocol (BGP)~\cite{rekhter2005border}, advertising (\ie originating) the prefixes that are allocated to them and adopting routes to reach prefixes that are originated by other ASes. BGP is a policy-based~\cite{gao2001stable}, destination-oriented path-vector protocol, where an AS receives paths to a destination prefix from its neighbors, selects the ``best'' path to prefer based on its local routing policies and other criteria~\cite{cisco2016bgp}, and advertises it to other neighbors based on its export policies.

\myitem{BGP (in-)security today.} While BGP is scalable and expressive in terms of routing policies, the lack of security by design (\eg authentication of IP prefix advertisements) is a critical limitation that frequently results in routing attacks, such as BGP hijacks. These attacks, based on the manipulation of routing tables via fraudulent advertisements of prefixes and/or AS-paths, have pestered the Internet for decades~\cite{youtube-hijack-2008,bitcoin-hijack-2014,backconnect-hijack-2016,iranian-hijack-2017,russian-hijack-2017,google-japan-hijack-2017,google-china-hijack-2018}.
To remedy this, several approaches have been proposed~\cite{mitseva2018state,sermpezis2018artemis}; the most prominent ones are (i) RPKI~\cite{bush2013resource}, to secure prefix-to-origin mappings, and (ii) BGPsec~\cite{lepinski2017bgpsec}, to cryptographically protect each hop of the AS-path to a prefix. However, securing inter-domain routing is an extremely slow process~\cite{goldberg2014taking}. While BGPsec is still practically not deployed~\cite{goldberg2011secure}, there are serious efforts on expanding RPKI adoption, spearheaded by large ISPs~\cite{atnt2019rpki} and IXPs~\cite{decix2019rpki}. Despite the rapid increase in RPKI Route Origin Authorizations (ROAs), and the adoption of Route Origin Validation (ROV)~\cite{nist2019rpki}, 
the highly centralized nature of RPKI w.r.t. PKI certificate authorities~\cite{cooper2013risk}, and its inability to protect networks against sophisticated hijacks (\eg AS-path manipulation more than 1 hop away from the legal origin of a prefix)
require complementary approaches.

\myitem{Blockchains and Applications.} A \textit{Blockchain}~\cite{nakamoto2008bitcoin} is a sequence of cryptographically protected and linked units of information, called blocks, and serves as a distributed, replicated, append-only, tamper-resistant ledger. Anyone can easily retrace the history of the transactions that are stored in the blocks; transactions are non-reputable. Numerous applications of this concept~\cite{swan2015blockchain} have been implemented mainly in the area of e-finance (\eg Bitcoin cryptocurrencies~\cite{nakamoto2008bitcoin}, Ethereum smart contracts~\cite{crosby2016blockchain}). Blockchain-based proposals have recently appeared in the context of managing and securing Internet resources, such as ASNs~\cite{xing2018bgpcoin}, IP prefixes~\cite{paillisse2018ipchain,xing2018bgpcoin}, BGP updates~\cite{hari2016internet,saadroutechain} and DNS records~\cite{liu2018data}. While these approaches employ Blockchain mechanisms to address the lack of security in the Internet, 
they are based on constructs that are complex to manage, scale up, and secure (\eg smart contracts~\cite{atzei2017survey}). This makes their application extremely challenging, hindering potential adoption~\cite{hogewoning2018}.

\myitem{Objectives.} In this work, our intention is to build a passive Blockchain-based system for storing and validating transactions related to IP prefix allocations, revocations and updates, as well as BGP path announcements and withdrawals. Drawing from the numerous lessons that the bumpy road towards adoption of RPKI and BGPsec has taught the networking community~\cite{cohen2016jumpstarting}, we focus on simplicity of design to enable the use of Blockchains as passive hijack detection systems.
Our approach should be compatible with previous efforts
that, e.g., form complementary chains for concurrent path validation~\cite{saadroutechain}, or produce genesis blocks for IP-to-AS assignments based on IANA-RIR-ISP-AS interactions~\cite{xing2018bgpcoin}.

\myitem{Contributions.} Our contributions in the context of Block\-chain systems for securing BGP~\cite{xing2018bgpcoin,saadroutechain,paillisse2018ipchain,hari2016internet} are:
\begin{enumerate}[topsep=0pt,itemsep=-1ex,partopsep=1ex,parsep=1ex]

\item Based on first principles~\cite{li2004first}, we design a simple, scalable and robust system that (i) works in parallel to existing inter-domain routing, (ii) can detect violations in both IP allocation actions and path advertisements, and (iii) can be incrementally deployed, surpassing the limitations of classic proactive security approaches~\cite{bush2013resource,lepinski2017bgpsec}.

\item We provide an open-source implementation that can be tested in the real world with minimal requirements. To the best of our knowledge, this is the first prototype that implements the concepts of the motivational position paper of Hari \etal~\cite{hari2016internet}; however, we do not use Blockchains to change BGP, but to validate its information.

\end{enumerate}
 
\myitem{Overview.} Section~\ref{sec:network} describes the formation of the overlay Blockchain network, and the (inter-)actions of the participating nodes.
Section~\ref{sec:chain} gives an overview of the basic entities of the Blockchain, \ie blocks and transactions, while Sections~\ref{sec:chain:transactions:ip-allocation} and~\ref{sec:chain:transactions:bgp-update} describe in detail the transactions specific to the resources that we want to secure (\ie IP prefixes and AS-paths, respectively), including their respective state and validation mechanisms.
Section~\ref{sec:impl} presents preliminary results using a minimal viable prototype, focusing on its scalability. After discussing related work in Section~\ref{sec:related}, we conclude the paper.

\vspace{-2mm}
\section{The Blockchain Network}\label{sec:network}
Here, we describe the basic operations that the networked
Blockchain instances (nodes-ASes) need to perform.

\myitem{Nodes.} In our Blockchain network, each node corresponds to an AS and runs its own Blockchain instance. It possesses: (i) an IP address and port, where the node is reachable, (ii) the ASN of the network where the node is running, and (iii) a pair of public and private keys for cryptographic operations.
Nodes form an overlay peer-to-peer full-mesh
~network, independently from BGP, and communicate over the Internet.

\myitem{Bootstrapping.} An initial group of nodes, called \emph{bootstrap nodes}, start populating the chain, using the IP prefix allocation Genesis block as the first block of the chain (see Section~\ref{sec:chain}).
These nodes serve as rendezvous points between new arrivals and already connected nodes. Every new node peers with them and requests their neighbors; therefore, they have a full view of the network at any given time.
Example bootstrap nodes could be well-connected ISPs or IXPs.

\myitem{Peering and Authenticating.} To become peers, two nodes need to exchange their public keys and associate them with node identifiers (ASNs). To scale this up, after discovering the rest of the network, new nodes broadcast their public keys together with their digitally signed information (IP address, port, ASN), and request the same data from their new peers. Key-to-ASN associations can be optionally mediated by IRs.
Peers periodically check the liveness of each other via ``keep-alive'' messages; non-responding peers are disconnected (from their neighbors' perspective).

\myitem{Making Transactions.} Nodes generate transactions according to their activity w.r.t. inter-domain routing; \eg they can allocate IP prefixes or advertise paths to these prefixes. Every new transaction is broadcasted to the network; every node is notified about and can validate incoming transactions, rejecting or logging invalid ones (\eg for real-time detection of routing attacks or offline forensics).

\myitem{Mining Blocks.} In our setup, any node can act as a miner; nodes can collect incoming and self-generated transactions and include them in blocks. Before starting the mining process, the miner must ensure that it has the most up-to-date --valid-- version of the chain and that the transactions it is about to mine are valid and do not already exist in the chain. Different schemes (we select \emph{Proof of Work-PoW}~\cite{nakamoto2008bitcoin}; \emph{Proof of Stake-PoS}~\cite{king2012ppcoin} is an alternative) can be used for ``proving'' the miner to the other nodes. After the mining is complete, the miner sends a message to all its neighbors to let them know that the chain has just been updated with one new block; they can in turn request the new chain from the miner.

\myitem{Validating Chains.} Whenever a node receives an --updated-- chain learned from another node, it needs to validate the chain's consistency and correctness before proceeding to the conflict resolution stage (\eg in case the new chain is different from its own). For each block, it (i) checks that the hash of the previous block is consistent, (ii) verifies the Proof of Work/Stake, and (iii) authenticates the miner of the block using the miner's public key and signature. 
Finally, the node can check whether the mined transactions in the chain are actually valid, using chain-derived states (see Sections~\ref{sec:chain:transactions:ip-allocation},~\ref{sec:chain:transactions:bgp-update}).

\myitem{Resolving Conflicts.} After validating a new chain, a node needs to decide if it should adopt it as the most recent --valid-- version. This decision is made using a conflict-resolution algorithm; we opt for the simple ``longest chain'' rule~\cite{nakamoto2008bitcoin}.

\vspace{-2mm}
\section{Blocks and Transactions}\label{sec:chain}
The Blockchain is the primary data structure that is replicated among the nodes of the Blockchain overlay network (see Section~\ref{sec:network}). It is a simply connected list of blocks, that are mined by these nodes. Each block contains one or more transactions associated with allocating IP prefixes to ASes and advertising (partial) AS-paths towards these prefixes. 
For the basic Blockchain design, we build on the required principles proposed in the original Bitcoin paper~\cite{nakamoto2008bitcoin}.

\myitem{Basic Blocks.} Each block includes the following data:
\begin{enumerate}[topsep=0pt,itemsep=-1ex,partopsep=1ex,parsep=1ex]
\item \texttt{transactions}: a list of all the transactions in this block.
\item \texttt{hash}: the hash of the block's content (\eg with a specified amount of leading zeros~\cite{nakamoto2008bitcoin} determining PoW difficulty).
\item \texttt{nonce}: the --tunable-- value used for the generation of the PoW of this block.
\item \texttt{previous hash}: the hash of the previous block in the chain, used to validate the consistency of the chain as an uninterrupted sequence of linked blocks.
\item \texttt{signature}: the digital signature of the block, signed by its miner using its private key.
\item \texttt{index}: the position of the block in the chain.
\item \texttt{timestamp}: the time when the block was built.
\item \texttt{mined\_timestamp}: the time when the block was mined.
\item \texttt{miner}: the ID (ASN) of the node that mined this block.
\end{enumerate}

\myitem{Genesis Blocks.} The first block to be included in the chain is called the \emph{Genesis Block}; this does not need to be mined, since the data in this block originate from reliable sources and are used to ``bootstrap'' the chain. Since we aim to protect IP prefixes and the paths leading to them, \emph{Genesis Blocks} include initial mappings of IP prefixes to ASNs, before networks start advertising them to the rest of the Internet. These data can stem from IRs~\cite{riperir2019} and can be pre-populated using the currently available RPKI ROAs~\cite{miro2019rpki} as ``ground truth''.

\myitem{Transactions.} Transactions are included in Blocks. Each transaction maps input data to output data, and includes:
\begin{enumerate}[topsep=0pt,itemsep=-1ex,partopsep=1ex,parsep=1ex]
    \item \texttt{input}: a list of input parameters and values, including what are the involved entities (sender, recipient), resources and corresponding constraints/rules. Example: ``ASX leases prefix P to ASY for a period of two months.''
    \item \texttt{output}: a list of output parameters and values, representing the result of this transaction, \eg ``Prefix P is now allocated to ASY for a period of two months''. 
    \item \texttt{type}: the type of the transaction, \ie:
    \texttt{IP Allocation - (Genesis) Assign/Update/Revoke} ~(see Section~\ref{sec:chain:transactions:ip-allocation}),
    \texttt{BGP Path - Announce/Withdraw} (see Section~\ref{sec:chain:transactions:bgp-update}). 
    \item \texttt{signature}: the digital signature of the transaction, signed by the node that created it,
    using its private key.
    \item \texttt{timestamp}: the time the transaction was made.
    \item \texttt{txid}: the unique ID of the transaction. 
\end{enumerate}

\myitem{Workflow.} Before proceeding to the description of specific transaction types, we need to understand what actions an AS needs to be able to perform in the context of inter-domain routing. We consider the following motivational workflow.
\begin{enumerate}[topsep=0pt,itemsep=-1ex,partopsep=1ex,parsep=1ex]
\item An IR assigns a prefix P to ASX with a specific lease period. We consider this a prefix ``allocation'' transaction from an AS to one or more ASes. Since IANA and other IRs can participate in such allocations, special ASNs can be used as their identifiers (in input/output parameters).
\item ASX allocates this prefix to organization Y, to be used by the siblings ASW and ASZ of Y for a specific time period.
\item ASW and ASZ advertise this prefix to their neighbors, which in turn propagate this announcement to their own neighbors. To encode this information, we do not need to include the entire AS-path vector in a transaction; the knowledge that ``this AS learned this prefix from these ASes and advertises it to those ASes'' is sufficient to build valid AS-level graphs leveraging transaction sequences.
\item ASW and/or ASZ withdraw the prefix since their lease period is about to expire or a network failure occurred.
\item ASX revokes the prefix allocation, reclaiming ownership.
\end{enumerate}
We classify transactions in \emph{IP Allocation} (Section~\ref{sec:chain:transactions:ip-allocation}) and \emph{BGP Path} (Section~\ref{sec:chain:transactions:bgp-update}) transactions.
By going through the transactions that are encoded --in sequence-- within the Blockchain, nodes/ASes recreate and update the corresponding states. Using these states, they validate transactions by comparing what they get with what they expect; if any transaction fails one of the checks, it is invalid and it is dropped and logged.

\vspace{-2mm}
\section{IP Allocation Transactions}\label{sec:chain:transactions:ip-allocation}
These transactions relate to the management and validation of the mapping between IP prefixes and ASNs.
The basic principle is enabling ASes to allocate/assign IP prefixes to other ASes, updating the period of this lease, or revoking it. 
Advocating simplicity of design,
our proposed \emph{Assign} transaction type can serve equivalently to the \emph{IP register}, \emph{IP allocate}, \emph{IP assign} and \emph{ROA add} transactions described in~\cite{xing2018bgpcoin}, 
reducing the overhead of complex resource state machines. Moreover, we consider transactions as simple input/output processes, avoiding coin trading and management implications~\cite{paillisse2018ipchain,xing2018bgpcoin}.


\myitemit{Genesis Assign.} This is the first transaction to be included in the chain (in the \emph{Genesis Block}). It starts either empty (for typical bootstrapping) or retrieves IP allocation data from trustworthy sources (including RPKI ROAs and IRs) and formats them. The input is the data retrieved from these sources (e.g., a ROA associating prefix P to ASX). The output is a list of ``ground truth'' \texttt{(Prefix, ASN)} allocations.

\myitemit{Assign.} This transaction is made when an AS (or IR) wants to allocate/assign the prefix it currently owns to one or more ASes (or IRs) for a specified duration of time (\ie lease). 
The original owner of this prefix loses any claim over it until the lease duration has expired. The input is a list of:
\begin{enumerate}[topsep=0pt,itemsep=-1ex,partopsep=1ex,parsep=1ex]
\item \texttt{prefix}: the prefix to be allocated.
\item \texttt{as\_source}: the AS (original prefix owner) that allocates the prefix and makes (and signs) the \emph{Assign} transaction.
\item \texttt{as\_dest}: the AS(es) to which the prefix is allocated.
\item \texttt{source\_lease}: the original lease duration as it was set for the prefix owner that makes this transaction.
\item \texttt{lease\_duration}: the lease duration ($\leq$ \texttt{source\_lease}).
\item \texttt{transfer\_tag}: flag determining sub-leasing capabilities.
\item \texttt{last\_assign}: the \texttt{txid} of the transaction that allocated the prefix to its original owner, \ie \texttt{as\_source}.
\end{enumerate}
The output is a list of updated IP-to-AS ownership entries, using one entry per destination (\ie recipient) ASN:
\texttt{(prefix, as\_dest, lease\_duration, transfer\_tag)}.

\myitemit{Update.} This transaction is made when an AS wants to update the --not expired-- lease period of a prefix it had previously assigned to another AS. The input is the following:
\begin{enumerate}[topsep=0pt,itemsep=-1ex,partopsep=1ex,parsep=1ex]
\item \texttt{as\_source}: the AS that updates the lease period and makes (and signs) the \emph{Update} transaction.
\item \texttt{assign\_tran}: \texttt{txid} of the original \emph{Assign} transaction.
\item \texttt{new\_lease}: new lease ($\leq$ \texttt{source\_lease(assign\_tran)})
\end{enumerate}
The output is a list of updated IP-to-AS ownership entries, using one entry per destination ASN of the initial \texttt{assign\_tran}:
\texttt{(prefix, as\_dest, new\_lease, transfer\_tag)}.

\myitemit{Revoke.} This transaction is made when an AS wants to reclaim the prefix from all the ASes it had previously assigned it to, and is only valid once the designated lease duration has actually expired. Revocations can be generated automatically. The input is a list of the following elements:
\begin{enumerate}[topsep=0pt,itemsep=-1ex,partopsep=1ex,parsep=1ex]
\item \texttt{as\_source}: the AS that reclaims the prefix and makes (and signs) the \emph{Revoke} transaction.
\item \texttt{assign\_tran}: \texttt{txid} of the revoked \emph{Assign} transaction.
\end{enumerate}
The output is: \texttt{(prefix, as\_source, transfer\_tag,\\new\_lease\_duration)}, where the first three elements are extracted directly from the original \emph{Assign} transaction, while the new lease is calculated as: \texttt{source\_lease(assign\_tran)} - \texttt{lease\_duration(assign\_tran)}.


\myitem{State.} The \emph{IP Allocation} state is collected at each node and holds information about the ownership of each IP prefix; which AS currently owns it, for how long, if it can transfer it to other ASes and the ID of the last related valid \emph{Assign} transaction.
~Initially, this state is populated using IP allocation data gathered from reliable sources, contained in the \emph{Genesis block} and encoded in a \emph{Genesis Assign} transaction. It gets updated whenever a new \emph{IP Allocation} transaction makes it into the chain. 
For \emph{Assign} transactions, nodes replace the current owner of a prefix with the ASes the owner is assigning the prefix to. For \emph{Revoke} transactions, nodes 
restore the original prefix owner. For \emph{Update} transactions, they simply update the lease periods of the respective allocation entries.

\myitem{Validating \emph{Assign} transactions.} First, the node needs to check whether \texttt{as\_source} is able to assign this prefix to \texttt{as\_dest}; this is feasible only if (i) \texttt{as\_source} currently owns the prefix, (ii) it can provide the \texttt{txid} of the last valid assignment that transferred the prefix to it (\texttt{last\_assign}), (ii) its \texttt{source\_lease} is greater than (or equal to) the new\\\texttt{lease\_duration}, and (iii) it has the right to transfer the prefix to others (\texttt{transfer\_tag}). It also checks if \texttt{as\_dest} are in the blockchain network and thus, verifiable.

\myitem{Validating \emph{Update} transactions.} The node first checks\\whether the lease of the current prefix owner(s) has not already expired. It then checks if the AS that makes this transaction (\texttt{as\_source}) is the one that made the original \emph{Assign} transaction (\texttt{assign\_tran}). Finally, it checks if the \texttt{as\_dest} in \texttt{assign\_tran}, currently own(s) the prefix. 

\myitem{Validating \emph{Revoke} transactions.} The node ensures that the lease has expired for revocation to be feasible. If it has expired, it executes the same checks as for the \emph{Update} transaction (\texttt{as\_source}, \texttt{assign\_tran} and \texttt{as\_dest} checks).

\section{BGP Path Transactions}\label{sec:chain:transactions:bgp-update}
These transactions relate to the management and validation of end-to-end AS-level paths towards IP prefixes, originated by ASes compliant to valid IP Allocations (Section~\ref{sec:chain:transactions:ip-allocation}). The basic principle is enabling ASes to advertise prefixes they learn from one or more ASes, to other ASes, as well as withdraw them. Valid paths towards prefixes can be then built by examining these ``learn-advertise'' operations in sequence. Note that by taking advantage of sequences of \emph{Announce} transactions of the form ``ASX learned prefix P from ASY and advertised it to ASW, ASZ'', instead of ``ASX has an AS-path AP towards prefix P'', we minimize the complexity of maintaining highly dynamic information for end-to-end AS-paths (\eg as encoded in BGP updates during a BGP path exploration process, or in complex structures of AS sub-groups as proposed in~\cite{saadroutechain}). Since we encode stable connectivity and policy information from the perspective of the AS that conducts the transaction, we save resources (see Section~\ref{sec:impl}).


\myitemit{Announce.} This transaction is made when an AS wants to advertise a prefix it learned from some --neighbor-- ASes to some other --neighbor-- ASes. Routing policies, such as valley-free policies~\cite{gao2001stable}, can be encoded both in the set of neighbors from which the prefix is learned (import policy) and the set of neighbors to which it is advertised (local and export policy). For privacy purposes, an AS can select itself what information it wants to publish in the Blockchain. The input of this transaction is a list of the following elements:
\begin{enumerate}[topsep=0pt,itemsep=-1ex,partopsep=1ex,parsep=1ex]
\item \texttt{prefix}: the prefix to be announced/advertised.
\item \texttt{as\_source}: the AS that advertises the prefix and makes (and signs) the \emph{Announce} transaction.
\item \texttt{as\_source\_list}: ASNs of neighbor ASes from which the prefix was learned.
\item \texttt{as\_dest\_list}: ASNs of neighbor ASes to which the prefix was advertised.
\end{enumerate}
The output is a list of partial paths towards the prefix, formed between the ASes of \texttt{as\_source\_list}, \texttt{as\_source}, and \\\texttt{as\_dest\_list}. For example, if \texttt{prefix} = $P$, \texttt{as\_source} = ASX, \texttt{as\_source\_list} = [AS1, AS2], and \texttt{as\_dest\_list} = [AS3, AS4], the output encodes the path set \{($P$-AS1-ASX-AS3), ($P$-AS1-ASX-AS4), ($P$-AS2-ASX-AS3), ($P$-AS2-ASX-AS4)\}. 

\myitemit{Withdraw.} This transaction is made when an AS wants to withdraw a specific prefix from all the ASes it had previously advertised it to, \eg due to a network failure or for policy-related reasons. The input is the following: 
\begin{enumerate}[topsep=0pt,itemsep=-1ex,partopsep=1ex,parsep=1ex]
\item \texttt{prefix}: the prefix to be withdrawn.
\item \texttt{as\_source}: the AS that withdraws the prefix and makes (and signs) the \emph{Withdraw} transaction.
\end{enumerate}
The output encodes the same information as the input.


\myitem{State.} The \emph{BGP Path} state is encoded in directed AS-level graphs per prefix. Similar to the \emph{IP Allocation} state, it is initially populated using the data of the \emph{Genesis Block}. First, a graph is created for every prefix; the first node of each graph is the prefix itself. After that, an edge between an AS node and the prefix node is added for every AS that owns the prefix based on the IP allocation data. After a new --valid-- \emph{Assign} transaction is inserted in the chain, the previous topology for this prefix is cleared, since the prefix owner(s) (and thus, the valid origin(s)) change; new edges between the ASes that are the new owners of the prefix, and the prefix node, are added. For \emph{Revoke} transactions, this process is reversed. \emph{Update} transactions do not affect this state. After a new \emph{Announce} transaction is added to the chain, new edges, from \texttt{as\_source} to \texttt{as\_dest\_list}, are added to the prefix graph. These edges are directed towards the prefix, following the traffic direction.
Finally, \emph{Withdraw} transactions result in the removal of the edges between the withdrawing node and its predecessors. All other nodes (and their corresponding edges) that cannot reach the prefix in the new regime are also removed. For simplicity, we assume that withdrawals precede revocations, meaning that the graph is already clear of stale routing information upon a \emph{Revoke} transaction.

\myitem{Validating \emph{Announce} transactions.} First, the node verifies the origin of this transaction (\texttt{as\_source}) based on the topology (AS-level graph) associated to this prefix; if \texttt{as\_source} does not have valid paths to the prefix via its \texttt{as\_source\_list} neighbors (from which it learned the prefix), the transaction is rejected. It then checks if all the ASes in \texttt{as\_source\_list} and \texttt{as\_dest\_list} are in the Blockchain network. Finally, it checks if this transaction introduces loops in the topology of this prefix; if it does, it is rejected.

\myitem{Validating \emph{Withdraw} transactions.}
Similarly to \emph{Announce} transaction checks, the node checks if there is at least one valid path from the withdrawing AS towards the prefix.


\vspace{-2mm}
\section{Preliminary Evaluation}\label{sec:impl}
\myitem{Basics.} We have implemented an open-source~\cite{sfirakis2019github} working prototype of the proposed Block\-chain in Python. 
Each running node instance has an IP address, port and ASN associated to it. Public keys are self-signed.
~Nodes peer with each other, form an overlay network, and execute the actions described in Section~\ref{sec:network} (\eg discovering neighbors, exchanging keys, making transactions, mining blocks, etc.) using a REST interface. For example, they issue \texttt{GET} requests to acquire the version of the chain from other peers, or \texttt{POST} requests to broadcast transactions they make to the network.

\myitem{Correctness.} To verify the correctness of our implementation, we test hypothetical chains with real BGP data. \emph{Genesis Assign} transactions are bootstrapped using the \texttt{pfx2as} dataset (IP prefix to ASN mappings) from CAIDA~\cite{caida2019pfx2as}, based on the entries of the Routing Information Bases (RIBs) of RouteViews~\cite{routeviews2019} BGP monitors. To form new \emph{BGP Path - Announce} transactions, we replay BGP updates, collected via BGPStream~\cite{bgpstream2019}, as seen on RouteViews and RIPE RIS~\cite{riperis2019} monitors,
and translate them by extracting AS-triplets (\ie [previous\_AS, advertising\_AS, next\_AS])]) from the AS-path included in the BGP update message. We verify that the AS-level graph for a prefix, as created via the raw BGP update data, is equal to the graph derived from processing the transactions of the chained blocks. We also check that valid transactions of any type have the desired effect on the Blockchain and its associated \emph{IP Allocation} and \emph{BGP Path} states. Moreover, 
we generate and broadcast invalid transactions, such as transactions with a fake creator, \emph{Assign/Update/Revoke} transactions from invalid origins, or \emph{Announce/Withdraw} transactions from nodes that do not have valid paths to the prefix(es) they want to announce/withdraw. We verify that such transactions are logged and discarded (even if they make it to the chain of a malicious or misconfigured node).
Note that transactions involving ASes off the network are not verifiable and are not added to the chain; however, as the Blockchain network scales up, more prefixes and AS-paths (on per-prefix AS-level graphs) can be successfully validated. 

\myitem{State Storage Requirements.} Next, we use back of the envelope calculations to estimate the storage capacity (in terms of RAM or HDD space) that the proposed Blockchain would require, per instance, in order to store the needed \emph{IP Allocation} and \emph{BGP Path} transactions and state. We focus on the storage ``heavy hitters''.
~The calculations are based on (i) the BGP report by Huston~\cite{huston2017bgp}, and (ii) AS-level topological statistics by CAIDA~\cite{caida2019topo}. We apply the notation of \textit{order of magnitude} for the calculations, defined as: $M(x) = \lfloor{log_{10}x}\rfloor,x\in\mathbb{R}$. For example, the number $80=8*10$ corresponds to an order of magnitude $M(80)=M(10)=1$. In our Python-specific implementation~\cite{sfirakis2019github}, we calculate the memory requirements both for IPv4 and IPv6-related transactions, for all types, as $M(10)$ bytes per transaction.
~W.r.t. \emph{IP Allocation} state, we store information worth of $M(10)$ bytes per prefix. According to~\cite{huston2017bgp}, $M(100k)$ IPv4 and $M(10k)$ IPv6 prefixes are currently advertised in BGP by $M(10k)$ ASes; $M(1)$ bytes are required for storing a single IPv4 prefix, $M(10)$ bytes for IPv6 and $M(1)$ bytes for an ASN. Therefore, we need $M(100k*1*10k*10+10k*10*10k*10)=M(10^{10})$ bytes, or $M(10)$ GBytes for maintaining the entire \emph{IP Allocation} state. This estimation is compliant ($M(10)$ GBytes for $\sim800K$ prefixes) with the empirically measured Blockchain size ($\sim1$ GByte for $\sim150k$ IP prefixes) found in~\cite{paillisse2018ipchain}; while we follow a different approach and architecture focusing on simplicity of design, our calculations correctly approximate the order of magnitude needed to store this kind of state. Following a similar logic for the per-prefix AS-level graphs (\emph{BGP Path} state), and taking into account that besides the $M(100k)$ IPv4 prefixes, $M(10k)$ IPv6 prefixes and $M(10k)$ ASNs we see $M(100k)$ AS-level links in BGP, we need $M(100k*1*100k*1+10k*10*100k*1)=M(10^{10})$ or $M(10)$ GBytes to store all AS-level prefix graphs.

To understand how fast the previous state increases, we consider the \emph{BGP Path} transactions, since the related information changes orders of magnitude faster than IP prefix ownership (which relies on slow management processes instead of dynamic topology and policy changes). 
According to~\cite{huston2017bgp}, BGP sees $M(100k)$ BGP updates per day. Each AS involved in an update (announcement/withdrawal) learned by one of its neighbors, propagates this update to its other neighbors. In the Blockchain, these updates sent to each neighbor are grouped in transactions containing several AS-triplets (\eg
~``ASX learned prefix P from ASA, ASB and advertised it to ASC, ASD''). 
An AS-path of length $L$ is broken into $L$ AS-triplets and thus $L$ such transactions; the average AS-path length in the Internet is $M(1)$~\cite{huston2017bgp}. Thus, the transaction (vs BGP update) counts are scaled down by a factor approximated by the neighbor degree of each AS (equal to $M(10)$, on average~\cite{caida2019topo}), and scaled up by the average path length; the Blockchain should see $M(1*100k/10)=M(10k)$ transactions per day related to BGP updates. Requiring $M(10)$ bytes per transaction, this translates to $M(10*10k)=M(100k)$ bytes in total per day. Our calculations are conservative, not accounting for savings in duplicate transaction counts vs BGP update bursts during the BGP path exploration process. 
 
 \begin{table}
\centering
\caption{Mining times for different PoW implementations.}\label{table:mining-speed}
\vspace{-3mm}
\begin{small}
\begin{tabular}{|c|c|c|c|c|c|}
\hline
{\# lead 0's} & {min(s)} & {max(s)} & avg(s) & med(s) & $90^{th}$ perc.(s)\\
\hline
4 & 0.0 & 1.4   & 0.2 & 0.9 & 1.5\\
5 & 0.1 & 12.8  & 3.1 & 3.6 & 8\\
6 & 0.3 & 223.8 & 54.7 & 31.6 & 160\\
\hline
\end{tabular}
\end{small}
\vspace{-5mm}
\end{table}

\myitem{Mining speed.} We use a server with average features (4 CPU cores, 16GB RAM) as a miner for \emph{BGP Path - Announce} transactions. By replaying raw BGP updates as \emph{Announce} transactions, we evaluate the time elapsed between the creation (and broadcasting) of a transaction, and the successful mining of the block that contains it. This time is indicative of the ``speed'' of the mining process, and thus whether the chain formation can keep up with online management of transactions. This is critical for Internet Blockchains to avoid the buffering of enormous backlogs (\eg BGP updates) during periods of bursty control-plane traffic. Our results for different difficulties of PoW are presented in Table~\ref{table:mining-speed}. Note that an increase in 1 leading zero leads to approximately a $10x$ increase in computation time. This demonstrates a trade-off between the load of incoming transactions to be mined and the robustness of the Blockchain. Larger loads require more transactions within a block, and faster mining times to avoid ever-increasing backlogs. However, speed comes at the price of robustness, since it may be easier for attackers to gather the necessary PoW power to corrupt the chain. 
Detection-wise, the Blockchain system can operate in real-time, since incoming transactions are validated the moment they are created and broadcasted to the network.

\section{Related Work}\label{sec:related}
Addressing the root cause of routing attacks, \ie the assumption of trust among ASes without proper authentication mechanisms~\cite{matsumoto2017authentication}, is a relatively old --but still unmet-- objective; for example, Chang \etal~\cite{chang2011trust} propose a behavioral assumption scheme that probabilistically computes AS reputation based on previous feedback and expectations. With the advent of Blockchains as systems that can facilitate transactions among mutually distrustful entities, without the need for trusted mediators, several related efforts have been done to secure Internet resources.
~Our work is influenced by the position paper of Hari \etal~\cite{hari2016internet}, who first introduced the concept of applying Blockchains in the context of IP prefixes and BGP updates; however, they focus more on the concepts rather than actual design and implementation. Xing et al.~\cite{xing2018bgpcoin} propose \emph{BGPcoin}; a Blockchain-based Internet number resource management system that leverages Ethereum smart contracts to protect IP prefixes and authenticate valid prefix origins, without though securing the end-to-end AS-path. 
Pailisse \etal~\cite{paillisse2018ipchain} follow a similar approach, based on PoS mining, to also validate IP prefix allocation/delegation, emphasizing in performance and scalability assessment. Based on a similar motivation as our work, Saad \etal~\cite{saadroutechain} propose \emph{RouteChain}, a system that employs Blockchains to validate BGP path advertisements. However, \emph{RouteChain} requires the dynamic formation of AS hierarchies and subgroups for quick consensus; this has been proven to be extremely challenging in practice~\cite{feamster2004some}.
Finally, there have been attempts to provide Blockchain-based Internet naming validation and/or secure DNS, such as \emph{Namecoin}~\cite{kalodner2015empirical}, \emph{Blockstack}~\cite{ali2016blockstack}, and \emph{DecDNS}~\cite{liu2018data}.


\vspace{-2mm}
\section{Conclusions}\label{sec:conclusions}
Following a first-principles approach for encoding inter-domain routing primitives, we designed and implemented an open-source prototype of a Blockchain that stores and validates transactions related to IP prefixes (assignments and revocations) and BGP paths towards these prefixes (announcements and withdrawals). We conducted a preliminary evaluation of the prototype and we showed that it can be hosted on servers with average technical capabilities (\eg storage capacity), 
being able to cope with the estimated load of transactions in today's Internet. Moreover, it can work --passively-- in concert with RPKI and other detection systems (based on Blockchain or not), \eg receiving ground truth prefix-to-AS data from RPKI ROAs or feeding hijack detectors with logged invalid \emph{BGP Path} transactions. 
Future research directions include:
(i) prototype deployment in at least two --neighboring-- ASes (e.g., IXP peers), (ii) investigation of PoW vs PoS, (iii) extensions of transaction data (e.g., BGP communities), (iv) mitigation of impact of flapping routes on \emph{BGP Path} transactions,
and (v) neighbor-selective \emph{BGP Path - Withdrawals} and \emph{IP Allocations}.

\section*{Acknowledgements}\label{sec:acks}
This work has been funded by the European Research
Council grant agreement no. 790575 (PHILOS project).

\bibliographystyle{ACM-Reference-Format}
\bibliography{references}

\end{document}